\documentclass[fleqn,10pt]{wlscirep}
\usepackage{soul}
\usepackage{hyperref}
\usepackage[normalem]{ulem}

\title{Towards Ultrafast Gyroscopes Employing Real-time Intensity and Spectral Domain Measurements of Ultrashort Pulses}

\author[1,2,*]{Maria Chernysheva}
\author[2]{Srikanth Sugavanam}
\author[2,3]{Sergei K. Turitsyn}
\affil[1]{Leibniz Institute of Photonic Technology, Albert-Einstein srt. 9, Jena 07745, Germany}
\affil[2]{Aston Institute of Photonic Technologies, Aston University, Aston Triangle, Birmingham B4 7ET, United Kingdom}
\affil[3]{Novosibirsk State University, 2 Pirogova Str., Novosibirsk, 630090, Russia}

\affil[*]{maria.chernysheva@leibniz-ipht.de}

\begin{abstract}
\textbf{ Active ring laser gyroscopes (RLG) operating on the principle of the optical Sagnac effect are preferred instruments for a range of applications, such as inertial guidance systems, seismology, and geodesy, that require both high bias stability and high angular velocity resolutions. Operating at such accuracy levels demands special precautions like dithering or multi-mode operation to eliminate frequency lock-in or similar effects introduced due to synchronisation of counter-propagating channels. Recently proposed bidirectional ultrafast fibre lasers can circumvent the limitations of continuous wave RLGs. However, their performance is limited due to the nature of the highly-averaged interrogation of the Sagnac effect. In general, the performance of current optical gyroscopes relies on the available measurement methods used for extracting the signal. Here, by changing the paradigm of traditional measurement and applying spatio-temporal intensity processing, we demonstrate that the bidirectional ultrafast laser can be transformed to an ultrafast gyroscope with acquisition rates of the order of the laser repetition rate, making them at least two orders of magnitude faster than commercially deployed versions. We also show the proof-of-principle for dead-band-free round trip time-resolved spectral domain measurements using the Dispersive Fourier Transform, further enhancing the gyroscopic sensitivity. Our results reveal the high potential of application of novel methods of signal measurements in mid-sized ultrafast fibre laser gyroscopes to achieve performances that are currently available only with large-scale RLGs.}

\end{abstract}
\begin{document}

\flushbottom
\maketitle

\thispagestyle{empty}

Improving the accuracy of relative positioning or rotational sensing is important both for fundamental science and for various practical engineering applications.  High-precision optical measurements have the potential to unlock new breakthrough methods and approaches in this field. With progress in the general understanding of optical phenomena in a laser cavity and the development of advanced laser configurations came the ability to measure ultraslow angular velocities. Thus, optical gyroscopes employing the Sagnac effect make it possible to detect rotations of the ground with 10$^{-11}$~rad$\cdot$s$^{-1}$ sensitivity, with an integration time of several hundreds of seconds\cite{belfi2017,Igel2007}, which enables observation of Chandler and Annual wobbles\cite{Schreiber2011}. Alternatively, for applications requiring fast data acquisition rates, such as when a gyroscope is a key part of inertial measurements unit for self-navigation\cite{prikhodko2018,escobar2018}, their substantially lower sensitivity is compensated for by high acquisition rates of up to several kilohertz\cite{Lefevre}.

Large-ring laser gyroscope technology is capable of providing highly sensitive inertial rotation measurements.
Among the impressive recent applications, one can mention the direct observation of the rotational microseismic noise \cite{hadziioannou2012} and the detection of very long period geodetic effects
on the Earth's rotation vector \cite{Schreiber2011}.  However, their application has both practical and fundamental limits and restrictions caused by their size, elaborate fabrication, and maintenance, and more importantly by the impact of the frequency lock-in effect. Backscattered light enhances the coupling between counter-propagating beams, causing carrier frequency synchronisation\cite{Chao1984lock,Loukianov1999}. As a result, the beat note signal disappears for a range of small angular velocities\cite{Schreiber2013}. Numerous approaches have been suggested to mitigate this limitation by decreasing backscattering. This includes the application of high-reflective dielectric coatings of cavity mirrors ($\sim$99.998\%), improvement of laser cavity geometry, or dithering the resonance frequency of the cavity with reference to an external laser beam\cite{Schreiber2013,Loukianov1999}. Owing to the lock-in effect, maintenance-free all-fibre configurations, which are typically considered beneficial, become disadvantageous for laser gyroscopes, as they suffer from Rayleigh scattering\cite{hansen1998rayleigh}. 

Alternative attempts for frequency lock-in effect elimination have exploited ultrashort pulses instead of continuous wave radiation in laser gyroscopes\cite{Fork1981,Salin1990}, since the counter-propagating ultrashort pulses interact only in two points in the cavity \cite{Buholz1967,Chesnoy1989,Krylov2017}. In such configurations, a differential phase shift of counter-propagating pulses due to the gyroscopic effect is generally evaluated by RF measurements of the beat-note frequency shift, that is, the change in the interference pattern of two frequency combs, corresponding to counter-propagating ultrashort pulses at each round trip\cite{Hendrie:16}.

To date, the interrogation of the Sagnac effect has relied on the study of its cumulative effect over time scales much larger than the typical cavity lifetime. Thus, existing methodologies impose a bottleneck on the gyroscope bandwidth, limiting it to a few kilohertz. The availability of ultrafast detectors and high-resolution real-time oscilloscopes has ushered in a multitude of novel, real-time methodologies for studying fibre laser dynamics in both the intensity and spectral domain. For instance, the method of spatio-temporal dynamics \cite{Churkin2015,Sugavanam2016} makes it possible to identify of specific features of interest in the laser output and to observe their evolution with round trip time resolution over several hundreds or even thousands of round trips \cite{Tarasov:15,Peng2016,dudley2014instabilities, Peng2018}. The Dispersive Fourier Transform (DFT)\cite{Goda2013} is a real-time spectral method that exploits the principle of Fraunhofer diffraction in the temporal domain and can be used to obtain the time-resolved spectra of successive mode-locked pulses \cite{Runge2013,Runge:15,Herink2016}. Such methods used to study the fast dynamics of fibre lasers also open up perspectives for interrogating the Sagnac effect in real-time, enhancing the functionality of gyroscopes. 

In this paper, we introduce a new concept of gyroscopic effect evaluation by analysing the dynamics of a pair of counter-propagating ultrashort soliton pulses applying three real-time measurement techniques in as a proof-of-principle demonstration bidirectional ultrafast fibre laser. Firstly, we show how the real-time spatio-temporal dynamics of the counter-propagating pulses can be used for direct observation and analysis of the temporal drift between solitons induced by the Sagnac effect accumulated over several round trips with a  resolution of $10^{-2}$ deg$\cdot$s$^{-1}$. Secondly, we show how such real-time measurements can be used to monitor angular velocities at a rate equivalent to the round trip time of the laser, resulting in effective gyroscope bandwidths that surpass currently available modalities by at least two orders of magnitude. Thirdly, we show how the synchronised regimes can be utilised for rotation sensing by employing real-time DFT measurements. All these techniques enable improvement of the sensitivity by at least an order of magnitude when compared to earlier demonstrated \cite{Krylov2017} by using mode-locked lasers, reaching $10^{-3}$ -- $10^{-4}$~deg$\cdot$s$^{-1}$. Simple theory based on a linear-regime approximation is presented, which helps to estimate the order of magnitude of sensitivities and errors and ascertain appropriate laser parameters needed to achieve the requisite resolution.

\section*{Results}

\begin{figure}[!b]
  \centering
\includegraphics[width=\textwidth]{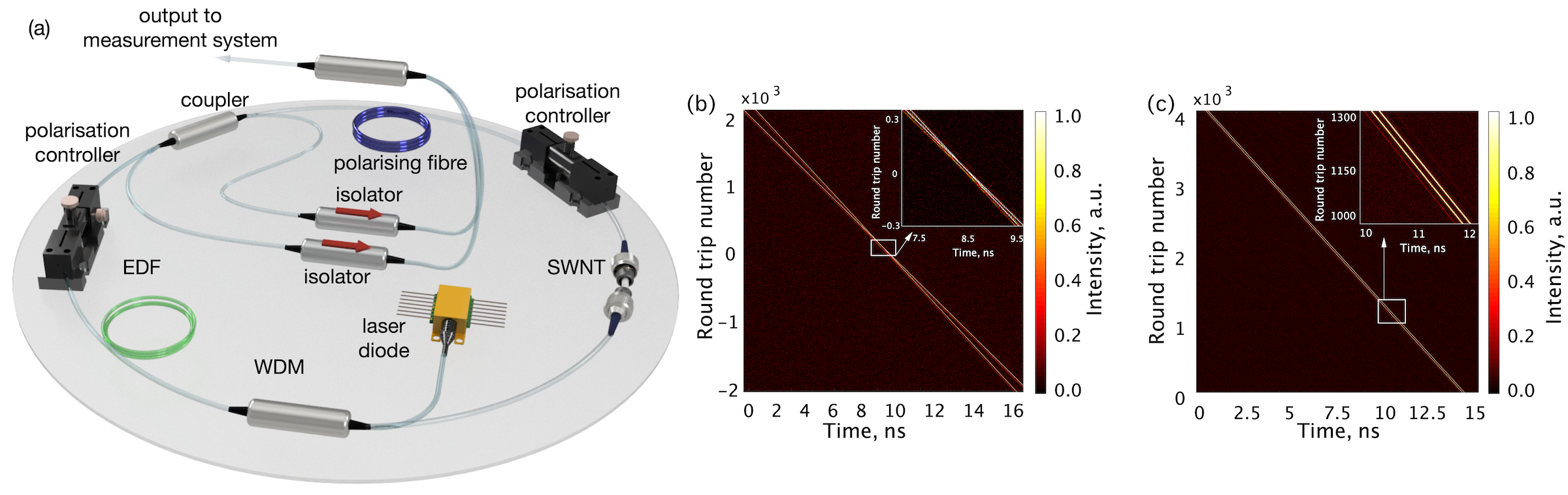}
\caption{\textbf{The ultrafast bi-directional laser and its operational regimes. }\textbf{a}, Schematic setup of the ultrafast fibre laser gyroscope. \textbf{b,c}, Spatio-temporal dynamics of pulses with different (\textbf{b}) and synchronised repetition frequencies (\textbf{c})}
\label{setup}
\end{figure}

\paragraph*{Our experiments} employ the hybrid mode-locked erbium-doped fibre laser setup (see \textit{Methods} section), placed on a rotating circular platform with a diameter of 0.62~m (Fig.~\ref{setup}a). The angular velocities of the platform with an ultrafast laser can be varied from 0 to 0.3 deg$\cdot$s$^{-1}$.  The stability of the ultrafast generation is ensured by a hybrid mechanism of passive mode-locking realised via single-walled carbon nanotube (SWNT) polymer saturable absorber and nonlinear polarisation evolution (NPE). NPE relies on the section of polarising fibre with bow-tie geometry and allows tuning of the nonlinear transfer function, predefining the nature of interactions between counter-propagating pulses. The laser cavity comprises of a 3-dB output coupler, which rotates with the entire interferometer cavity, creating the difference in the optical paths for counter-propagating pulses as reported for other simple Sagnac interferometers \cite{arditty1981sagnac}. Afterwards, the counter-propagating pulses are combined via a 3dB coupler, the output of which was used for the real-time measurements.

The experiment showed that two stable bidirectional regimes are possible: when repetition rates of counter-propagating pulses differ by tens of hertz (Fig.~\ref{setup}b), and when pulses in both channels are generated at the same repetition rate of 14.78~MHz (Fig.~\ref{setup}c).  In the latter case, the length of ports of the coupler, combining counter-propagating pulses, were adjusted to achieve pulse separation of $\sim$100~ps. We use different techniques for each of these two laser generation regimes to evaluate the gyroscopic effect. The case of channels with different repetition rate has been analysed by using spatio-temporal dynamics\cite{Churkin2015,Sugavanam2016real} (see \textit{Supplementary note 1}), and synchronised channels were measured with the DFT technique (see \textit{Supplementary note 2}).

\paragraph*{Time-domain analysis}
The methodology of spatio-temporal dynamics -- that is, a two-dimensional representation of laser intensity evolution over round trips -- make it possible to observe round trip time-resolved dynamics of laser pulses (see \textit{Supplementary note S1}). Figure~\ref{setup}b shows the spatio-temporal dynamics of the combined laser output over 20~000 consecutive round trips, for the platform at rest. When combined, the separation between pulses changes in the course of round trips; pulses overlap in the interim. The evolution traces for both pulses present straight lines, indicating that there are no attractive or repulsive forces between counter-propagating pulses. 

\begin{figure}[!t]
  \centering
\includegraphics[width=0.9\textwidth]{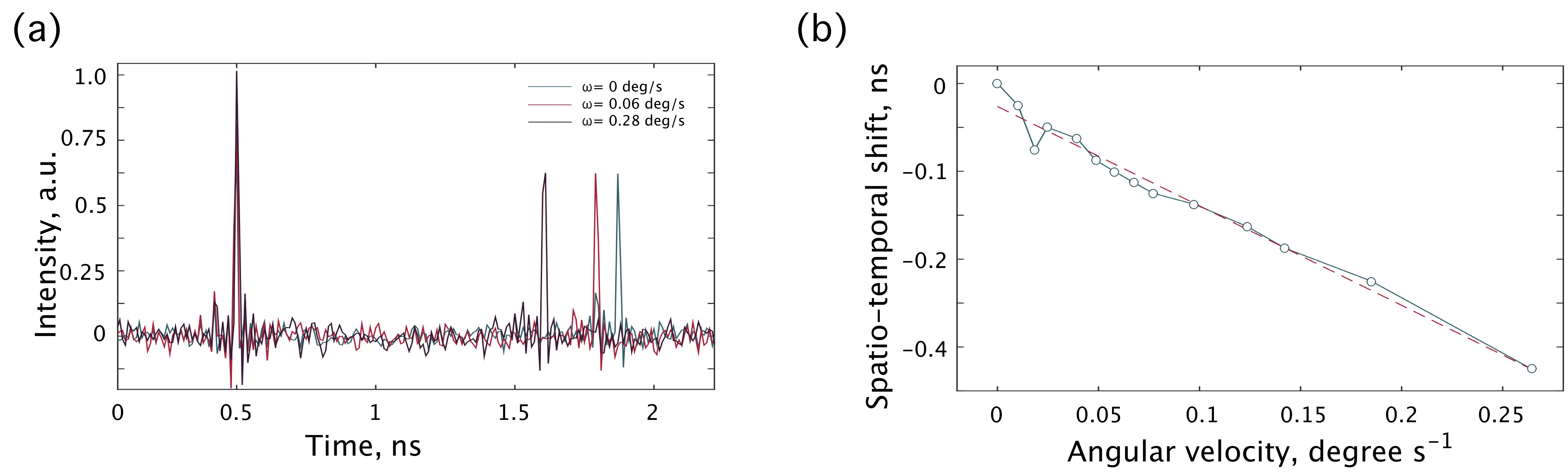}
\caption{\textbf{Observing the Sagnac effect in the time domain.} \textbf{a} Temporal shift of combined pulses at three different angular velocities. \textbf{b} Pulse temporal shift due to Sagnac effect (in reference to the platform in rest) at 10~000 round trips beyond point of pulse overlapping.}
\label{mov}
\end{figure}

Figure~\ref{mov}a shows the relative temporal separation between the CW and CCW pulses after $10^{4}$ round trips at different rotation velocities of the platform, showing that the rotation of the stage influences the cavity conditions for the pulses. The separation between the pulses obtained in the rest condition is the analogue of the conventional gyroscopic bias offset, and its effect can be removed from the actual measured values to obtain the drift introduced solely due to the Sagnac effect (Fig.~\ref{mov}b). This clearly shows the existence of a linear relation between the angular velocity and the corresponding relative pulse temporal drift. We attribute the deviations from the linear approximation to imperfections of the rotation stage, mainly owing to the slippage between the platform and the motor. The pulse separation is seen to decrease over round trips, owing to the direction of rotation of the stage. With the stage rotating in the opposite direction, the pulse separation would increase with round trips. In other words, the bidirectional laser operating in the non-synchronised regime can be used not only for ascertaining the magnitude of the angular velocity but also the direction of rotation.

The sensitivity $S$ of the gyroscope for this particular measurement configuration can be obtained directly from the slope of the linear fit in Fig.~\ref{mov}b, and is the analogue of the conventionally used gyroscopic scale factor. Here, it is estimated to be -1.13~ns/(deg$\cdot$s$^{-1}$)]$^{-1}$, or 0.885 deg.s$^{-1}$/ns. The primary uncertainty of measurement in this configuration can be attributed to the finite temporal resolution $\delta t_{res}$ (here, 25~ps) and can be estimated as $S\cdot\delta t_{res}$ to be 22.12~mdeg$\cdot$s$^{-1}$ (384~$\mu$rad$\cdot$s$^{-1}$). A theoretical estimate of the scale factor can be obtained from well-known expressions for the Sagnac effect \cite{arditty1981sagnac} (see Eq.~\ref{EqSpatTempSagnac} in \textit{Methods}), as 539~$\mu$deg$\cdot$s$^{-1}$ (9.4~$\mu$rad$\cdot$s$^{-1}$), which is much smaller than the experimentally obtained value. The differences can be attributed to the fact that the analytical expression Eq.~\ref{EqSpatTempSagnac} in \textit{Methods} only takes into account the linear effects as brought about by the Sagnac effect, not considering the nonlinear effects introduced across the laser cavity, similar to observed in nonlinear loop mirrors \cite{smith1995,gabitov1997}. Yet, the experimental results show there is a clear linear relationship between the angular velocity of the stage and the relative pulse separation, thus allowing us to use the current configuration for high accuracy angular velocity determination. Indeed, based on this understanding, and the nature of the real-time measurements made, it can be shown that the variation of the value of the scale factor, i.e. the bias stability, is of the order of 1.8E-10 ppm (see Supplementary note 4). The resolution limit can be countered for a given laser configuration by only increasing the number of round trips $N_{RT}$. Here, since the repetition period of the pulses in counter-propagating channels is $\sim$66~ns and the pulse relatively shift is 0.295~ps within one round trip time, one can consider at least 1.98$\times$10$^5$~round trips after pulses coincide (for the platform in rest). Thus, analysis of entire recorded dataset would make it possible to increase the resolution of the gyroscopic measurements by more than one order of magnitude.  

\begin{figure}[!b]
  \centering
\includegraphics[width=0.95\textwidth]{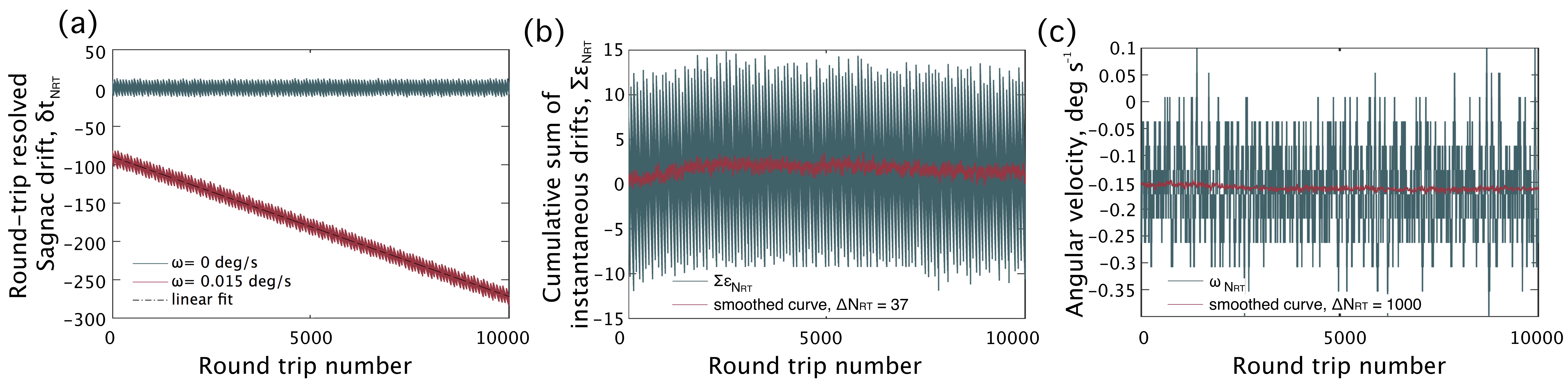}
\caption{\textbf{round trip time-resolved gyroscopic measurements.} \textbf{a}, Temporal separation between CW and CCW pulses introduced by the Sagnac effect, for the stage at rest (blue curve) and in motion (orange curve). The drift due to the gyroscopic bias has been compensated. Black dashed line is a linear fit, indicative of constant angular velocity. \textbf{b}, Blue curve - Residual of the linear fit function, which is a cumulative sum $\Sigma\epsilon_{N_{RT}}$ of the instantaneous deviations of the pulse separation over each round trip. The jagged appearance is a sampling artefact, which has been removed by a smoothing operation to give the orange curve. \textbf{c}, round trip resolved angular velocity obtained using Eq.\ref{RealTime}, where $\epsilon_{N_{RT}}$ is obtained from a first-order difference of the smoothed curve in Fig. b. The orange curve here is again obtained by a smoothing operation to remove the sampling artefacts}
\label{realtime}
\end{figure}

\paragraph*{Ultrafast gyroscope} In the above, the cumulative Sagnac time shift accrued after $10^4$ round trips was used to estimate the average angular velocity of the stage over the equivalent time period. One can obtain a measure of the angular velocity when $N_{RT}=1$ (see Eq.~\ref{EqSpatTempSagnac} in \textit{Methods}). While this method decreases the gyroscope sensitivity by a factor of $N_{RT}$, it has the potential to increase the effective gyroscope bandwidth to the order of the cavity repetition frequency, making it at least three orders of magnitude faster than commercially available gyroscopes, and, to the best of our knowledge, one of the highest bandwidth gyroscopes, even surpassing MEMs technologies. 

Figure~\ref{realtime}a shows the round trip resolved temporal separation $\delta T_{N_{RT}}$ introduced between the CW and CCW pulses. The gyroscopic bias offset has been removed, leaving behind only the Sagnac effect induced temporal drift between the pulses. The plot indicates a linearly increasing separation between the pulses (Fig.~\ref{realtime}a). As in Fig.~\ref{mov}b, the pulse separation is seen to decrease over round trips owing to the direction of rotation of the stage, which agrees with the definition in the Eq.~\ref{EqSpatTempSagnac} in \textit{Methods}. Thus, the average Sagnac temporal drift $\left<\delta t_{Sagnac} \right>$ over each round trip can be obtained from the slope of the linear fit function (Fig.~\ref{realtime}a, dotted line). Deviations about this linear fit, $\epsilon_{N_{RT}}$ can result from instantaneous drifts of the stage away from the mean angular velocity. Figure~\ref{realtime}b shows the residuals of the linear fit, which is not $\epsilon_{N_{RT}}$ but its cumulative sum up to the round trip $N_{RT}$ -- that is, $\sum_{i = 1}^{N_{RT}}\epsilon_{N_{RT}}$. $\epsilon_{N_{RT}}$ -- can thus be obtained a simple first-order difference of the residuals of the linear fit (Fig.~\ref{realtime}c, also see \textit{Supplementary note~6}). Thus, with knowledge of $S$ and the instantaneous temporal drifts, the round trip time-resolved angular velocity measurements as shown in Fig.~\ref{realtime}c (blue curve) can be obtained directly from time-domain spatio-temporal dynamics, via the formula:

\begin{equation}\label{RealTime}
\Omega_{N} = S\cdot\left[\left<\delta t_{Sagnac} \right> + \epsilon_{N}\right].
\end{equation}

For the current laser, in principle, the acquisition rate or gyroscope bandwidth can be as high as $418\cdot10^3$ samples per second (or 418~kHz), where the scale factor $S$ is 94~mrad$\cdot$s$^{-1}$, calculated using Eq.~\ref{EqSpatTempSagnac} in \textit{Methods}. This value is estimated under the consideration of a finite temporal resolution of the real-time oscilloscope and using the value of the scale factor obtained using the linear approximation (see \textit{Supplementary note 5}). However, as the scale factor in the experiment is drastically reduced owing to the nonlinearities, the bandwidth obtained in the experiment is closer to 19 kHz, which is an order of magnitude higher than current state-of-the-art and commercially available instruments \cite{Li17,liu2017high}. The finite bandwidth effect has been taken into account in the above instantaneous angular velocity dynamics by incorporating a moving-window smoothing operation (orange curve, Fig.~\ref{realtime}c). We would like to stress that the increase of the gyroscopic bandwidth always comes at the expense of the decrease of the gyroscope sensitivity by the same factor. Defining the final application of gyroscope will allow identification of suitable accuracy versus bandwidth trade off. While the angular velocity of our stage is currently limited to about 0.2 deg$\cdot$s$^{-1}$, the above methodology is not limited by the angular velocity, and actually offers better gyroscope bandwidths with increasing angular velocities.

\paragraph*{DFT analysis} 
Figure~\ref{setup}(c) shows the second regime of operation of the bidirectional laser, where the CW and CCW pulses have equal repetition rates, resulting in their observed parallel trajectories on the spatio-temporal dynamics. The length of output coupler fibre ports was chosen in such way to ensure small separation between counter-propagating pulses when combined. As can be seen in Fig.~\ref{setup}(c) or Fig.~S3 pulses preserve separation of $\sim$100~ps. The locking of the pulse repetition rates of counter-propagating pulse trains is caused by passive synchronisation owing to cross-phase modulation \cite{smith1995} and the SWNT dynamics \cite{Chow1986,Zhang2011,Yoshitomi2005}. The transition from the non-synchronised mode described above to the pulse separation locked regime does not modify the pulse parameters within experimental accuracy (see Fig.~\ref{laser_param}). 

\begin{figure}[!b]
  \centering
\includegraphics[width=\textwidth]{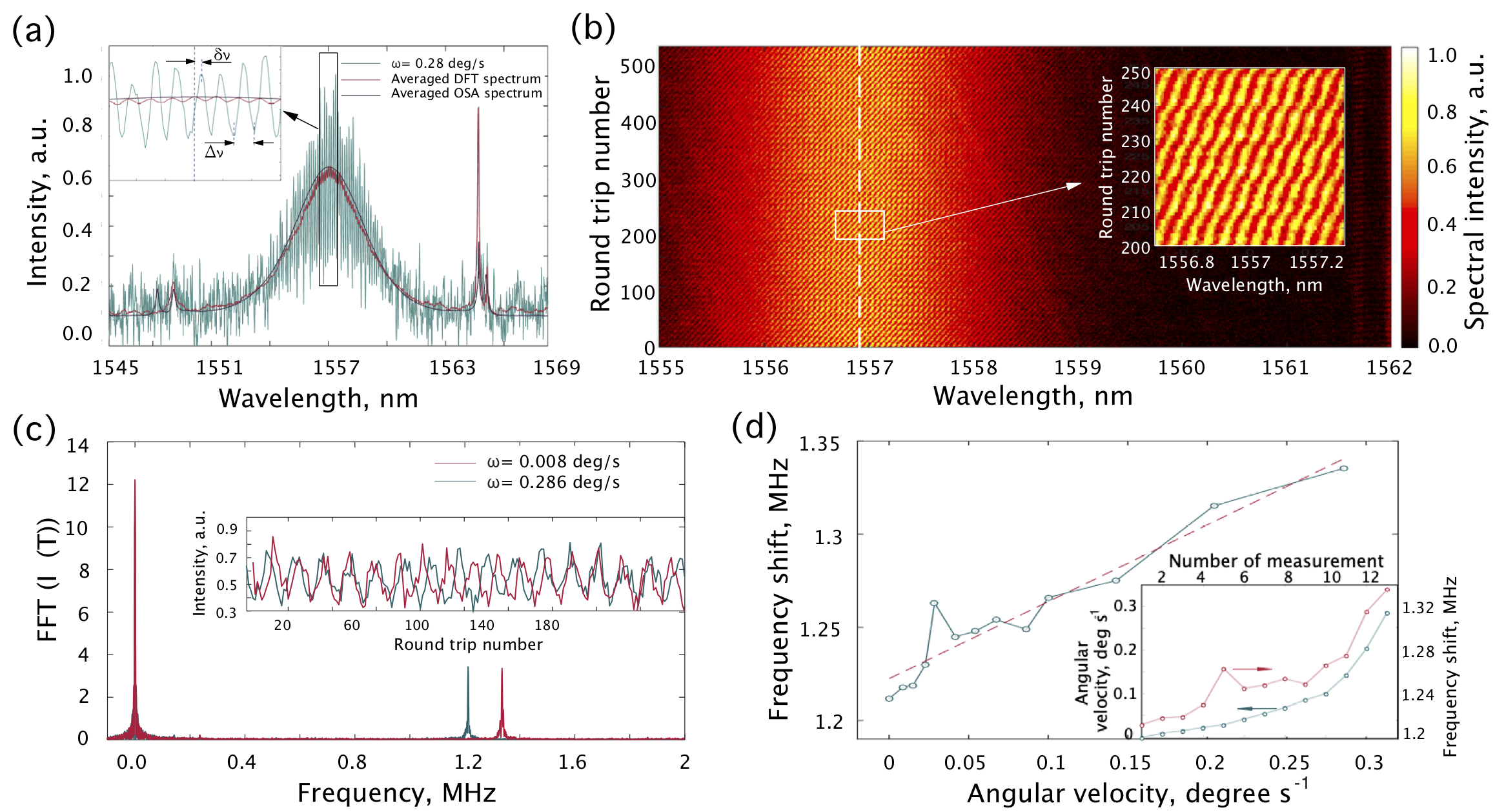}
\caption{\textbf{Gyroscopic measurements by single-shot spectral evolution.} \textbf{a}, Single-shot spectrum, averaged single-shot spectra over 5000 round trips and optical spectra, measured with OSA, of combined pulses. \textbf{b}, Single-shot spectra dynamics within 500 round trips and Inset: Zoomed-in central part of spectral evolution. \textbf{c}, Frequency shift for two different angular velocities. Inset: Spectral intensity fringing at $\sim$1556.9~nm -- along white dashed line in (b)}. \textbf{d}, DFT response to different angular velocities. Inset: Qualitative correlation of frequency shift and angular velocity during measurements.
\label{stab}
\end{figure}

When such pulse synchronisation occurs, the influence of the Sagnac effect cannot be directly observed by using intensity domain spatio-temporal approaches. To prove that the pulses exhibit stationary behaviour regardless gyroscope rotation, the dynamics of autocorrelation function over round trips was investigated (see \textit{Supplementary note S2}) for different platform rotation velocities. To analyse gyroscopic effect using this regime, we shift to the spectral domain and apply the DFT technique (see \textit{Supplementary note S2}). The single-shot spectrum of combined pulses is highly modulated due to interference between stretched combined pulses (Fig.~\ref{stab}a and its inset -  blue plot). It is known that the frequency of modulation of the interference pattern is inversely related to the temporal separation $t_{sep}$ of the pulses $\Delta\nu=1/t_{sep}$\cite{chin1992}. The averaged single-shot spectrum (red curve in Fig.~\ref{stab}a) is in good agreement with the time-averaged spectrum recorded with the optical spectrum analyser (OSA) for combined pulses (navy curve), the position of characteristic Kelly side-bands correlates with one in individual spectra of counter-propagating pulses. The wavelength axis is obtained by mapping the obtained spectrum with the averaged one measured using OSA. In the previous scenario, the gyroscopic effect introduces a change in the temporal separation, which should cause a change in the modulation frequency. Here, however, the temporal separation between the pulses does not change appreciably with changing angular velocity of the table (see Fig.~S3), rendering the change in the modulation frequency unresolvable. 

Here, we can utilise the additional time-scale available to us -- that is, the evolution of the spectrum over round trips -- to reveal the presence of the gyroscopic effect. Figure~\ref{stab}b demonstrates recorded single-shot spectra over 5~000 consecutive round trips. The gyroscopic effect in this regime of operation of the laser would manifest in the form of a change in the tilt of the modulation of the DFT spectrum (see inset in Fig.~\ref{stab}b), ascertained by measuring a change in the frequency of the spectrally resolved intensity variation over round trips at $\sim$1556.9~nm (white dashed line, Fig.~\ref{stab}b). The magnitude of the effect can be obtained by calculating the FFT of the spectrally resolved intensity dynamics. Figure~\ref{stab}c shows how the change in the tilt is converted to a change in the frequency of modulation $f_{mod}$, as a function of the angular velocity of the table. Here, the FFT is accounted for the dynamics measured over 5000 round trips. The inset in Fig.~\ref{stab}c shows the round trip spectral intensity evolution for two angular velocities, measured along the spectral maximum (dashed line in Fig.~\ref{stab}b). 

The inset in Fig.~\ref{stab}d shows a one-to-one correspondence between the trends of the actual angular velocity and the magnitude of the gyroscopic shift as revealed by the DFT modulation tilt, justifying the use of the linear fit (red line in Fig.~\ref{stab}d). The values obtained using DFT-based measurements were individually confirmed via beat note measurements (see \textit{Supplementary note S3}). The tilt in the spectral modulation when the platform is in rest appears due to initial differences in carrier frequencies of counter-propagating pulses, producing carrier-envelope offset (CEO). The slope of the linear fit can then determine the gyroscopic sensitivity $S_{DFT}$, here 17.2~mdeg$\cdot$s$^{-1}$/kHz, with an error of (3~kHz)$^{-1}$ (as set by the FFT resolution). The use of the DFT methodology allows resolution of the angular velocity of 7.2~mdeg$\cdot$s$^{-1}$ (125~$\mu$rad$\cdot$s$^{-1}$), while the theoretically predicted resolution according to Eq.~\ref{EqDFTSagnac} (see \textit{Methods}) when $N_{RT}$ = 5000 is 10~$\mu$deg$\cdot$s$^{-1}$. Here, the trade-off is a loss in temporal resolution of the angular velocity drift. The resolution can be enhanced via moving window FFTs or even higher-order windowing operations like the Wigner Ville distribution \cite{Boashash2015,Sugavanam2016real}. The sensitivity can be enhanced by observing the intensity evolution over longer periods, here up to 3.96$\times$10$^5$ round trips. Therefore, the sensitivity is in principle limited by the oscilloscope memory.

\paragraph*{Discussion and conclusion} 

Ours is the first demonstration of the application of real-time intensity and spectral domain approaches to measuring the gyroscopic effect in a bidirectional ring ultrafast fibre laser. A direct time domain measurement of the Sagnac effect was previously considered as a low-accuracy gyroscopic signal processing method. However, the advent of high-bandwidth detectors and oscilloscopes allow us to interrogate this effect directly by using the recently emerged methodologies of spatio-temporal dynamics and DFT. We have shown in proof-of-principle experiments how the spatio-temporal dynamics approach can increase gyroscope readout rates up to two orders of magnitude higher than commercially available options. The measurement configuration is also highly simplified, requiring only the use of a coupler for combining the counter-propagating pulses; accurate contemporisation of the pulses becomes less of an issue. For experimental gyroscopic effect evaluation, real-time spatio-temporal dynamics takes pulse relative motion as an advantage. The DFT approach also implies tolerable pulse separation. This technique eliminates carrier-to-envelope offset frequencies noise caused bias frequency drift \cite{Kim2016} and, hence, does not require additional stabilising elements to be introduced into a laser cavity. For conventional active configurations offering comparable resolutions, lasers used typically require active stabilisation from environmental effects. Here, however, no form of stabilisation or even thermal isolation was applied. This can be attributed in part to the relatively short time-scales investigated and to the high stability of the ultrafast laser (jitter $<$ 0.8 ps). The combination of ultrashort pulse fibre laser design, together with the demonstrated signal processing approaches, presents significantly new and highly promising techniques of interrogating the Sagnac effect. The capabilities of these techniques can be enhanced further, approaching specifications of large ring laser gyroscopes, by the improvement of direct detection electronic systems for signal analysis. The performance of modern optical gyroscopes is often limited not by intrinsic physical effects, but rather by the available measurement methodology. Recently emerged techniques for characterisation of optical field and laser radiation in particular \cite{Churkin2015,Sugavanam2016,dudley2014instabilities, Peng2018,Tarasov:15,Peng2016,Goda2013,Runge2013,Runge:15,Herink2016} has the potential to revolutionise the field of optical gyroscopes.

The current laser configuration provides a rotation resolution of  384~$\mu$rad$\cdot$s$^{-1}$ in the spatio-temporal approach for $N_{RT} = 10^4$  round trips and 125~$\mu$rad$\cdot$s$^{-1}$ for DFT analysis over 5~000 round trips. Within the available range of platform rotation velocities, none of the presented techniques has demonstrated dead-band in gyroscopic measurements. While more advanced and large-scale active laser configurations still offer better stability and accuracy, the results obtained using the relatively simple and compact bidirectional ultrafast laser hold a possibility to be further improved by more than an order of magnitude within the current laser design by increasing the number of round trips, or alternatively, by using a resonator of larger scale. Extension of the laser cavity can lead to highly non-trivial pulse dynamics, as indicated by the observed deviation of the behaviour of the studied laser from the linear Sagnac approximation. This was also confirmed in different context of achieving ultralong cavity operation \cite{kobtsev2008,kieu2008}. The general methods proposed here can be applied in various applications beyond gyroscopes, helping to reveal the underlying physics of bidirectional ultrafast lasers. Currently, none of existing theoretical model is capable to completely describe the dynamics of the interaction between counter-propagating solitons inside the bidirectional mode-locked laser, most critically in saturable absorber or their non-local interaction in the gain medium. Our further plans include development of such numerical model to analyse pulse dynamics in rotating cavity, which will be published elsewhere.

In this work, we used a bidirectional mode-locked laser as a convenient and straightforward platform to generate and study dynamics of ultrashort pulses during cavity rotation. However, recent works on seismology and gyroscopy using interferometers based on telecommunication optical fibre cables\cite{marra2018}, passive ultrafast FOGs \cite{Kieu2017} or microresonator gyroscopes \cite{silver2017} clearly demonstrate the high potential of enhancing their performance further by using the novel approaches and methodology of measurements demonstrated here. Although it remains a subject for further investigation, we anticipate that the availability of new a generation of measurement techniques will lead to development of new technology solutions for real-time gyroscopes.

\bibliography{sample}

\section*{Acknowledgements}

MC acknowledges the support of Royal Academy of Engineering and Global Challenges Research Fund. SKT acknowledges the support by the Russian Science Foundation (Grant No. 17-72-30006). The work of SS was supported by the project MULTIPLY.

\section*{Author contributions statement}

MC conceived the laser experiment. SS suggested the concept of DFT analysis and the ultrafast gyroscope. MC and SS performed the gyroscopic effect evaluation and modelling. SKT analysed the results and overall supervised the work. All authors discussed and analysed the results and contributed to the preparation of the manuscript.

\section*{Additional information}

\textbf{Competing Interests:} The authors declare no competing financial interests.

\textbf{Correspondence and requests} for materials should be addressed to M.C.
\newpage

\section*{Methods}

\subsection*{Experimental setup} 
\begin{figure}[h]
  \centering
\includegraphics[width=0.9\textwidth]{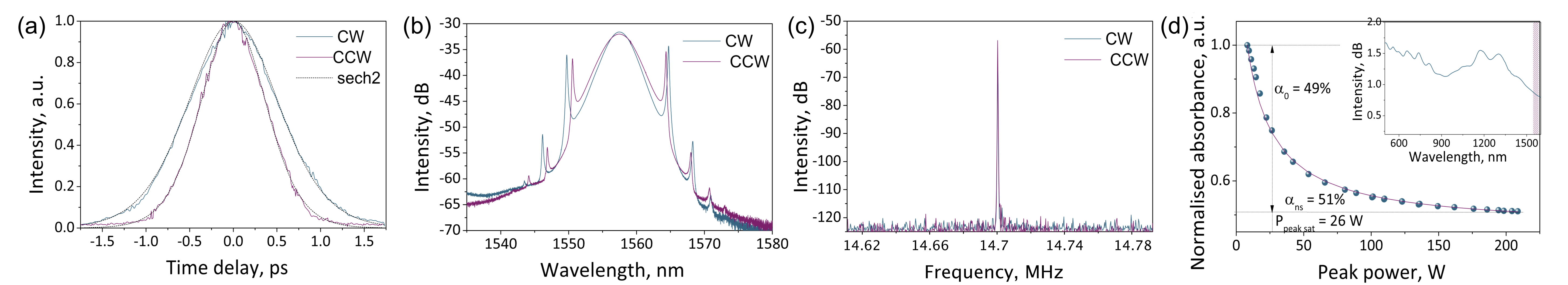}
\caption{\textbf{Laser output characteristics for both CW and CCW channels.} \textbf{a}, Autocorrelation traces. \textbf{b}, Output spectra. \textbf{c}, RF spectrum of the fundamental repetition rate. \textbf{d}, Saturation behaviour of SWNT. Inset: linear absorption spectrum of SWNT.}
\label{laser_param}
\end{figure}

In the experiment, the laser setup is similar to that presented in \cite{Chernysheva_bidir} (see Fig.~\ref{setup}). The fibre laser has a ring cavity configuration. A 2-m erbium-doped fibre (EDF) (Liekki Er30-4/125) is pumped via laser diode at 980 nm through 980/1550 wavelength division multiplexer (WDM). The total cavity length is 13.5 m with a corresponding fundamental repetition rate of $\sim$14.7~MHz (Fig.~\ref{laser_param}c). The ultrashort pulse formation is realised via hybrid mode-locking, that is, by using simultaneously nonlinear polarisation evolution (NPE) and single-walled carbon nanotubes (SWNT). The NPE, being a fast saturable absorber, is achieved by introducing 6-m coiled polarising (PZ) fibre (HB1550Z from Thorlabs), and a pair of polarisation controllers (PC), one of which is driven electronically (EPC). The bow-tie geometry of PZ fibre modifies the optical fibre refractive index and, therefore, creates different cut-off wavelength and attenuation for the orthogonally polarised fibre modes. The extinction ratio between slow and fast polarisation axis is $\sim$30~dB within the bandwidth of $\sim$130~nm around 1550~nm. The SWNT/polymer sample features high modulation depth around  54\%, non-saturable losses of 46\% and the saturation intensity of 58.8 MW~cm$^{-2}$ (Fig.~\ref{laser_param}d). The individual pulse parameters of each clockwise (CW) and counter-clockwise (CCW) channels are demonstrated in Fig.~\ref{laser_param} when the platform is in rest. The laser generates near transform-limited soliton pulses in both operation directions (Fig.~\ref{laser_param}a,b). The output power levels at the CW and CCW channels are around 1~mW, and pulse duration is 790 and 570 fs, respectively (Fig.~\ref{laser_param}a). These parameters have not changed with the laser platform rotation. Ultrashort pulses from both laser outputs passing through isolators are combined via a 3-dB coupler. 

The evaluation of gyroscopic effect via real-time dynamics is realised using a 33~GHz real-time oscilloscope (Agilent DSOX93204A) and a 50-GHz fast photodetector (Finisar XPDV2320R) with the minimum temporal resolution of the entire measurement system of $\sim$25~ps.

\subsection*{Gyroscope scale factor}
\paragraph{Spatio-temporal dynamics analysis} The difference in inclinations between the CW and CCW pulses (as in Fig.~\ref{setup}b) arises due to a difference in their round trip time, in turn resulting from a difference in their carrier frequencies. This intrinsic difference between the round trip times can be taken as an offset for our configuration and can be likened to the conventionally indicated gyroscopic bias offset value for conventional gyroscopes. When the stage starts to rotate with the angular velocity $\Omega$, the optical Sagnac effect leads to a further change in the cavity condition for the CW and CCW beams. The additional frequency difference introduced by the Sagnac effect can be recast as an increment to the cavity round trip time difference $\delta T_{N_{RT}}$ accrued after $N_{RT}$ round trips as:

\begin{equation}\label{EqSpatTempSagnac}
\delta T_{N_{RT}} = \frac{1}{f_{rep}^2}\frac{4A\Omega N_{RT}}{\lambda L} \equiv \frac{\Omega}{S}.
\end{equation}
here, $f_{rep}$ -- is the pulse repetition rate, $\lambda$ -- is the laser central wavelength, \textit{A} -- is area of the ring laser. Here, \textit{S} takes the notion of the scale factor for our bidirectional ultrafast laser-based gyroscope and specifies the sensitivity of laser gyroscope. For the laser currently employed, the sensitivity is 5.39~(deg$\cdot$s$^{-1}$)$\cdot$s$^{-1}$, where we have taken $N_{RT}=1$ and $\delta T_{N_{RT}}~=~25$ ps, which is equal to the temporal resolution limit of our measurement configuration $\delta t_{res}$. For a fixed laser configuration, the sensitivity can be further increased by observing the relative temporal shift between the CW and CCW pulses after several round trips (i.e., for $N_{RT}\gg1$).
When using storage oscilloscopes, the recorded number of round trips can be as large as $\sim$3.96$\cdot$10$^5$, thus increasing the sensitivity of the gyroscope by the same order. 

\paragraph{DFT analysis}
The sliding of the spectral fringes in the interference pattern of combined pulses (Fig.~\ref{stab}b) is caused by the infinitesimal drifting relative phase of combined pulses. Analogous to the above mentioned case, the tilt of modulation is referred to gyroscopic bias offset. The optical Sagnac effect introduces additional change in optical path lengths for counter-propagating pulses, causing further change in their carrier frequencies and, therefore, their relative phase when combined. In ultrafast bidirectional laser gyroscope cavity, the phase change due to Sagnac effect $\phi=8\pi A N_{RT}\Omega/(\lambda Lf_{rep})$ results in the change in CEO. Conventionally, the CEO frequency was analysed using beat-note measurements. The CEO of a complex of two pulses can be converted to a phase value as $\phi = 2\pi\delta\nu/\Delta\nu$, where $\Delta\nu$ -- is frequency of spectral modulation of interference pattern of individual single-shot spectrum and $\delta\nu$ -- instantaneous CEO. The frequency of modulation over consecutive round trips (along the white dashed line in Fig.~\ref{stab}b) $f_{mod}$ over $N_{RT}$ round trips can be estimated, analogous to Eq.~\ref{EqSpatTempSagnac} as:

\begin{equation}\label{EqDFTSagnac}
f_{mod}=\frac{f_{rep}\delta\nu}{\Delta\nu}=\frac{4AN_{RT}}{\lambda L f_{rep}}\Omega\equiv \frac{\Omega}{S_{DFT}}
\end{equation}
Here, \textit{R} is the radius of the gyroscopic platform with coiled laser setup, and $S_{DFT}$ is the scale factor of the ultrafast fibre laser gyroscope when analysed using DFT technique. With $N_{RT}$~=~1, the sensitivity -- that is, scale factor $S_{DFT}$ -- is 58~kHz/(deg$\cdot$s$^{-1}$).

\end{document}